# BODY-FIXED ORBIT-ATTITUDE HOVERING AT EQUILIBRIA NEAR AN ASTEROID USING NON-CANONICAL HAMILTONIAN STRUCTURE

Yue Wang[*] and Shijie Xu[†]

Orbit-attitude hovering of a spacecraft at the natural relative equilibria in the body-fixed frame of a uniformly rotating asteroid is discussed in the framework of the full spacecraft dynamics, in which the spacecraft is modeled as a rigid body with the gravitational orbit-attitude coupling. In this hovering model, both the position and attitude of the spacecraft are kept to be stationary in the asteroid body-fixed frame. A Hamiltonian structure-based feedback control law is proposed to stabilize the relative equilibria of the full dynamics to achieve the orbit-attitude hovering. The control law is consisted of two parts: potential shaping and energy dissipation. The potential shaping is to make the relative equilibrium a minimum of the modified Hamiltonian on the invariant manifold by modifying the potential artificially. With the energy-Casimir method, it is shown that the unstable relative equilibrium can always be stabilized in the Lyapunov sense by the potential shaping with sufficiently large feedback gains. Then the energy dissipation leads the motion to converge asymptotically to the minimum of the modified Hamiltonian on the invariant manifold, i.e., the relative equilibrium. The feasibility of the proposed stabilization control law is validated through numerical simulations in the case of a spacecraft orbiting around a small asteroid. The main advantage of the proposed hovering control law is that it is very simple and is easy to implement autonomously by the spacecraft with little computation. This advantage is attributed to the utilization of dynamical behaviors of the system in the control design.

**INTRODUCTION**

The great and increasing interest in the exploration of small primitive solar system bodies and the mitigation of near-Earth objects (NEOs) hazard has translated into an increasing number of small body missions. Detailed remote mapping, in-situ exploration and sampling of these primitive bodies can provide the scientists with unique and valuable information on the history and the evolution of our solar system. The major space agencies are all involved on missions to asteroids for the in-situ exploration or NEO hazard mitigation.

During the in-situ exploration of the asteroid and the asteroid deflection mission, the close proximity-operations are necessary. Generally, the motion of the spacecraft is very complicated due to the irregular mass distribution of the asteroid, the rotation of the asteroid, and the perturbations caused by the solar radiation pressure. Consequently, the orbital trajectories around such bodies are generally complex and non-periodic, and the stability of close orbits is guaranteed for a limited set of latitudes, as shown by Lara and Scheeres[1], and Scheeres[2][3].

---

[*] PhD Candidate, Department of Aerospace Engineering, School of Astronautics, Beihang University, Beijing, 100191, China. e-mail: ywang@sa.buaa.edu.cn
[†] Professor, Department of Aerospace Engineering, School of Astronautics, Beihang University, Beijing, 100191, China. e-mail: starsjxu@163.com

One strategy for the motion control proposed to meet these challenges is the hovering control [4][5][6][7][8]. During the hovering, the control force is used to null out the gravitational and rotational accelerations of the spacecraft, create and maintain an artificial equilibrium at a desired hovering position. This approach is feasible near small bodies because the gravitational field is weak and the nominal accelerations on a spacecraft are small. Unlike the traditional orbital control, hovering control can be used to explore over any point of the asteroid. The perturbing forces can also be actively nulled by the close-loop hovering controller. These are generally two types of hovering: hovering in the asteroid body-fixed frame and in the inertial frame. In the body-fixed hovering, the spacecraft is kept to a constant position in the asteroid body-fixed frame. The body-fixed hovering can be used to obtain high-resolution measurements and sample a particular area on the surface of the asteroid.

In the previous works on the hovering control mentioned above, the spacecraft has been modeled as a point mass, and the attitude control of the spacecraft was not considered. Actually, in many cases of the body-fixed hovering, the attitude of the spacecraft is also required to be kept stationary with respect to the asteroid. Moreover, the gravitational coupling between the orbital and rotational motions of the spacecraft can be significant in the close proximity of an asteroid due to the large ratio of the dimension of the spacecraft to the orbit radius, as shown by Koon et al.[9], Scheeres[10], Wang and Xu[11]. In Reference [12], the orbit-attitude hovering control of a spacecraft in the body-fixed frame of a uniformly rotating asteroid was designed using geometric mechanics, in which the gravitational orbit-attitude coupling of the spacecraft was considered. The body-fixed orbit-attitude hovering means that both the position and attitude of the spacecraft are kept to be stationary in the asteroid body-fixed frame.

The full dynamics, in which the spacecraft is modeled as rigid body, has been proposed to take into account the gravitational orbit-attitude coupling for a spacecraft moving in the proximity of an asteroid[13][14][15][16]. The non-canonical Hamiltonian structure of the problem in the proximity of an irregular-shaped asteroid has been uncovered in details by Wang and Xu[15][17]. The relative equilibrium is the stationary point of the Hamiltonian on the invariant manifold, which is the level set of the Casimir functions in the phase space. The nonlinear stability of the relative equilibrium is determined by the minimality of the Hamiltonian constrained on the invariant manifold[18]. Notice that the body-fixed orbit-attitude hovering is actually creating and maintaining an artificial relative equilibrium in the full spacecraft dynamics. In the present paper, we only consider the orbit-attitude hovering at a natural relative equilibrium. The hovering at an artificial relative equilibrium will be studied in the future.

During the close proximity-operations near an asteroid, the high level autonomy is a key element to overcome the long time delay in the communication from Earth and the complex environment. This requires that the control law has a simple form and needs little computation. The control design approach based on the non-canonical Hamiltonian structure is very promising in this aspect, since dynamical behaviors of the system can be fully utilized in this method.

A Hamiltonian structure-based feedback control law is proposed to stabilize the natural relative equilibria of the full dynamics to achieve the orbit-attitude hovering. This hovering control law is consisted of two parts: potential shaping and energy dissipation. The potential shaping is used to modify the gravitational potential artificially through position and attitude feedback so that the relative equilibrium is a minimum of the modified Hamiltonian on the invariant manifold.

The energy-Casimir method, which is originally used for determining nonlinear stability of a non-canonical Hamiltonian system, is used to obtain the condition of Lyapunov stability of the potential shaping feedback control. Then, the energy dissipation will lead the motion to converge asymptotically to the minimum of the modified Hamiltonian on the invariant manifold, i.e., the relative equilibrium. The feasibility of the proposed Hamiltonian structure-based stabilization feedback control law is validated through numerical simulations in the case of a spacecraft orbiting around a small asteroid.

**NON-CANONICAL HAMILTONIAN STRUCTURE AND RELATIVE EQUILIBRIA**

As described by Fig. 1, we consider a rigid spacecraft $B$ moving around a uniformly rotating asteroid $P$, the gravity field of which is approximated by a second degree and order-gravity field with harmonics $C_{20}$ and $C_{22}$. The mass center of the asteroid is assumed to be stationary in the inertial space, and the asteroid is rotating around its maximum-moment principal axis. The body-fixed reference frames of the asteroid and the spacecraft are given by $S_P=\{u, v, w\}$ and $S_B=\{i, j, k\}$ with $O$ and $C$ as their origins respectively. The origins of the body-fixed frames $S_P$ and $S_B$ are fixed at the mass center of the bodies, and the coordinate axes are chosen to be aligned along the principal moments of inertia.

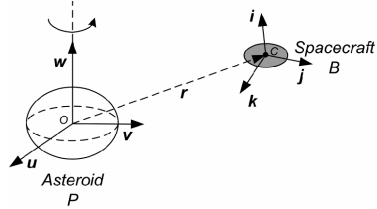

Figure 1. The rigid spacecraft moving around a uniformly rotating asteroid.

The coordinates of this non-canonical Hamiltonian system can be chosen as:

$$z = \left[ \boldsymbol{\Pi}^T, \boldsymbol{\alpha}^T, \boldsymbol{\beta}^T, \boldsymbol{\gamma}^T, \boldsymbol{R}^T, \boldsymbol{P}^T \right]^T \in \mathbb{R}^{18}, \tag{1}$$

where $\boldsymbol{\alpha}$, $\boldsymbol{\beta}$ and $\boldsymbol{\gamma}$ are coordinates of the unit vectors $u$, $v$ and $w$ expressed in the spacecraft body-fixed frame $S_B$; $\boldsymbol{\Pi}$, $\boldsymbol{R}$ and $\boldsymbol{P}$ are the angular momentum, position vector and linear momentum of the spacecraft respectively, which are all expressed in the body-fixed frame $S_B$[15][17]. The Poisson bracket $\{\cdot,\cdot\}_{\mathbb{R}^{18}}(z)$ of this non-canonical Hamiltonian system can be written as:

$$\{f, g\}_{\mathbb{R}^{18}}(z) = (\nabla_z f)^T \boldsymbol{B}(z)(\nabla_z g). \tag{2}$$

The Poisson tensor $\boldsymbol{B}(z)$ is given by[17]:

$$\boldsymbol{B}(z) = \begin{bmatrix} \hat{\boldsymbol{\Pi}} & \hat{\boldsymbol{\alpha}} & \hat{\boldsymbol{\beta}} & \hat{\boldsymbol{\gamma}} & \hat{\boldsymbol{R}} & \hat{\boldsymbol{P}} \\ \hat{\boldsymbol{\alpha}} & 0 & 0 & 0 & 0 & 0 \\ \hat{\boldsymbol{\beta}} & 0 & 0 & 0 & 0 & 0 \\ \hat{\boldsymbol{\gamma}} & 0 & 0 & 0 & 0 & 0 \\ \hat{\boldsymbol{R}} & 0 & 0 & 0 & 0 & \mathbf{I}_{3\times 3} \\ \hat{\boldsymbol{P}} & 0 & 0 & 0 & -\mathbf{I}_{3\times 3} & 0 \end{bmatrix}, \tag{3}$$

where $\mathbf{I}_{3\times 3}$ is the $3\times 3$ identity matrix, and the hat map $\wedge : \mathbb{R}^3 \to so(3)$ is the usual Lie algebra isomorphism. For a vector $w = [w^x, w^y, w^z]^T$,

$$\hat{w} = \begin{bmatrix} 0 & -w^z & w^y \\ w^z & 0 & -w^x \\ -w^y & w^x & 0 \end{bmatrix}. \tag{4}$$

The 18×18 Poisson tensor $B(z)$ has six geometric integrals as independent Casimir functions

$$C_1(z) = \boldsymbol{\alpha}^T\boldsymbol{\alpha}/2 \equiv 1/2, \quad C_2(z) = \boldsymbol{\beta}^T\boldsymbol{\beta}/2 \equiv 1/2, \quad C_3(z) = \boldsymbol{\gamma}^T\boldsymbol{\gamma}/2 \equiv 1/2,$$

$$C_4(z) = \boldsymbol{\alpha}^T\boldsymbol{\beta} \equiv 0, \quad C_5(z) = \boldsymbol{\alpha}^T\boldsymbol{\gamma} \equiv 0, \quad C_6(z) = \boldsymbol{\beta}^T\boldsymbol{\gamma} \equiv 0.$$

The twelve-dimensional invariant manifold is defined in $\mathbb{R}^{18}$ by Casimir functions:

$$\Sigma = \left\{ \left[ \boldsymbol{\Pi}^T, \boldsymbol{\alpha}^T, \boldsymbol{\beta}^T, \boldsymbol{\gamma}^T, \boldsymbol{R}^T, \boldsymbol{P}^T \right]^T \in \mathbb{R}^{18} \mid C_1(z) = C_2(z) = C_3(z) = \frac{1}{2}, C_4(z) = C_5(z) = C_6(z) = 0 \right\}. \tag{5}$$

The Hamiltonian of the system is given by[15]:

$$H(z) = \frac{1}{2}\boldsymbol{\Pi}^T \boldsymbol{I}^{-1} \boldsymbol{\Pi} + \frac{|\boldsymbol{P}|^2}{2m} - \omega_T \boldsymbol{\Pi}^T \boldsymbol{\gamma} - \omega_T \boldsymbol{P}^T (\hat{\boldsymbol{\gamma}} \boldsymbol{R}) + V(z), \tag{6}$$

where the diagonal matrix $\boldsymbol{I} = diag\{I_{xx}, I_{yy}, I_{zz}\}$ and $m$ are the inertia tensor and mass of the spacecraft respectively, $\omega_T$ is the angular velocity of the asteroid, and $V(z)$ is the gravitational potential. The second-order approximation of the potential $V(\boldsymbol{R}, \boldsymbol{\alpha}, \boldsymbol{\beta}, \boldsymbol{\gamma})$ is given by:

$$V(\boldsymbol{R}, \boldsymbol{\alpha}, \boldsymbol{\beta}, \boldsymbol{\gamma}) = -\frac{\mu m}{R} - \frac{\mu}{2R^3}\left[ tr(\boldsymbol{I}) - 3\bar{\boldsymbol{R}}^T \boldsymbol{I}\bar{\boldsymbol{R}} - m\tau_0 + 3m\tau_0 (\boldsymbol{\gamma} \cdot \bar{\boldsymbol{R}})^2 + 6m\tau_2 ((\boldsymbol{\alpha} \cdot \bar{\boldsymbol{R}})^2 - (\boldsymbol{\beta} \cdot \bar{\boldsymbol{R}})^2) \right], \tag{7}$$

where $\mu = GM$, $G$ is the Gravitational Constant, $M$ is the mass of the asteroid, $\tau_0 = a_e^2 C_{20}$, $\tau_2 = a_e^2 C_{22}$, $a_e$ is the mean equatorial radius of the asteroid, and $\bar{\boldsymbol{R}} = \boldsymbol{R}/|\boldsymbol{R}|$ [19].

The equations of motion can be written in the Hamiltonian form

$$\dot{z} = \boldsymbol{B}(z)\nabla_z H(z). \tag{8}$$

The explicit equations of motion can be obtained from Eqs. (6) and (8) as follows:

$$\begin{bmatrix} \dot{\boldsymbol{\Pi}} \\ \dot{\boldsymbol{\alpha}} \\ \dot{\boldsymbol{\beta}} \\ \dot{\boldsymbol{\gamma}} \\ \dot{\boldsymbol{R}} \\ \dot{\boldsymbol{P}} \end{bmatrix} = \boldsymbol{B}(z) \begin{bmatrix} \boldsymbol{I}^{-1}\boldsymbol{\Pi} - \omega_T \boldsymbol{\gamma} \\ \partial V/\partial \boldsymbol{\alpha} \\ \partial V/\partial \boldsymbol{\beta} \\ -\omega_T \boldsymbol{\Pi} - \omega_T \hat{\boldsymbol{R}}\boldsymbol{P} + \partial V/\partial \boldsymbol{\gamma} \\ -\omega_T \hat{\boldsymbol{P}}\boldsymbol{\gamma} + \partial V/\partial \boldsymbol{R} \\ -\omega_T \hat{\boldsymbol{\gamma}}\boldsymbol{R} + \boldsymbol{P}/m \end{bmatrix} = \begin{bmatrix} \hat{\boldsymbol{\Pi}}\, \boldsymbol{I}^{-1}\boldsymbol{\Pi} + \sum_{b=\boldsymbol{\alpha},\boldsymbol{\beta},\boldsymbol{\gamma},\boldsymbol{R}} \hat{\boldsymbol{b}}(\partial V/\partial \boldsymbol{b}) \\ \hat{\boldsymbol{\alpha}}(\boldsymbol{I}^{-1}\boldsymbol{\Pi} - \omega_T \boldsymbol{\gamma}) \\ \hat{\boldsymbol{\beta}}(\boldsymbol{I}^{-1}\boldsymbol{\Pi} - \omega_T \boldsymbol{\gamma}) \\ \hat{\boldsymbol{\gamma}}(\boldsymbol{I}^{-1}\boldsymbol{\Pi}) \\ \hat{\boldsymbol{R}}(\boldsymbol{I}^{-1}\boldsymbol{\Pi}) + \boldsymbol{P}/m \\ \hat{\boldsymbol{P}}(\boldsymbol{I}^{-1}\boldsymbol{\Pi}) - \partial V/\partial \boldsymbol{R} \end{bmatrix}. \tag{9}$$

Notice that $\sum_{b=\boldsymbol{\alpha},\boldsymbol{\beta},\boldsymbol{\gamma},\boldsymbol{R}} \hat{\boldsymbol{b}}(\partial V/\partial \boldsymbol{b})$ and $-\partial V/\partial \boldsymbol{R}$ are actually the gravity gradient torque and the gravitational force of the spacecraft expressed in the body-fixed frame $S_B$ respectively[17]:

$$\boldsymbol{T} = \boldsymbol{R} \times \frac{\partial V(\boldsymbol{R}, \boldsymbol{\alpha}, \boldsymbol{\beta}, \boldsymbol{\gamma})}{\partial \boldsymbol{R}} + \boldsymbol{\alpha} \times \frac{\partial V(\boldsymbol{R}, \boldsymbol{\alpha}, \boldsymbol{\beta}, \boldsymbol{\gamma})}{\partial \boldsymbol{\alpha}} + \boldsymbol{\beta} \times \frac{\partial V(\boldsymbol{R}, \boldsymbol{\alpha}, \boldsymbol{\beta}, \boldsymbol{\gamma})}{\partial \boldsymbol{\beta}} + \boldsymbol{\gamma} \times \frac{\partial V(\boldsymbol{R}, \boldsymbol{\alpha}, \boldsymbol{\beta}, \boldsymbol{\gamma})}{\partial \boldsymbol{\gamma}}, \tag{10}$$

$$F = -\frac{\partial V(\boldsymbol{R}, \boldsymbol{\alpha}, \boldsymbol{\beta}, \boldsymbol{\gamma})}{\partial \boldsymbol{R}}. \tag{11}$$

According to Reference [15], the relative equilibrium of the spacecraft corresponds to the stationary point of the Hamiltonian constrained by the Casimir functions. The stationary points can be determined by the first variation condition of the variational Lagrangian $\nabla F(z_e) = \boldsymbol{0}$, where

$$F(z) = H(z) - \sum_{i=1}^{6} \mu_i C_i(z). \tag{12}$$

By using Eq. (12), the equilibrium conditions are obtained as follows:

$$\boldsymbol{I}^{-1} \boldsymbol{\Pi}_e - \omega_T \boldsymbol{\gamma}_e = \boldsymbol{0}, \tag{13a}$$

$$-\frac{\mu}{2R_e^3}\left[12m\tau_2(\boldsymbol{\alpha}_e \cdot \overline{\boldsymbol{R}}_e)\overline{\boldsymbol{R}}_e\right] - \mu_1 \boldsymbol{\alpha}_e - \mu_4 \boldsymbol{\beta}_e - \mu_5 \boldsymbol{\gamma}_e = \boldsymbol{0}, \tag{13b}$$

$$\frac{\mu}{2R_e^3}\left[12m\tau_2(\boldsymbol{\beta}_e \cdot \overline{\boldsymbol{R}}_e)\overline{\boldsymbol{R}}_e\right] - \mu_2 \boldsymbol{\beta}_e - \mu_4 \boldsymbol{\alpha}_e - \mu_6 \boldsymbol{\gamma}_e = \boldsymbol{0}, \tag{13c}$$

$$-\omega_T \boldsymbol{\Pi}_e - \omega_T \hat{\boldsymbol{R}}_e \boldsymbol{P}_e - \frac{\mu}{2R_e^3}\left[6m\tau_0(\boldsymbol{\gamma}_e \cdot \overline{\boldsymbol{R}}_e)\overline{\boldsymbol{R}}_e\right] - \mu_3 \boldsymbol{\gamma}_e - \mu_5 \boldsymbol{\alpha}_e - \mu_6 \boldsymbol{\beta}_e = \boldsymbol{0}, \tag{13d}$$

$$-\omega_T \hat{\boldsymbol{P}}_e \boldsymbol{\gamma}_e + \left.\frac{\partial V}{\partial \boldsymbol{R}}\right|_e = \boldsymbol{0}, \tag{13e}$$

$$-\omega_T \hat{\boldsymbol{\gamma}}_e \boldsymbol{R}_e + \frac{\boldsymbol{P}_e}{m} = \boldsymbol{0}, \tag{13f}$$

where the subscript $e$ is used to denote the value at the equilibrium.

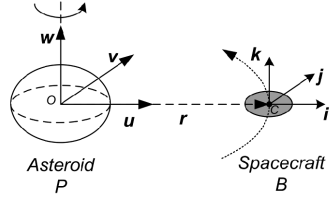

**Figure 2. One of the classical type of relative equilibria.**

We have given a classical type of relative equilibria of the spacecraft in Reference [15]. At this type of relative equilibria, the mass center of the spacecraft is located at the principal axes of the asteroid $u$ or $v$. The body-fixed frame $S_B$ is parallel to the body-fixed frame of the asteroid $S_P$, i.e., $\boldsymbol{\alpha}_e$, $\boldsymbol{\beta}_e$ and $\boldsymbol{\gamma}_e$ are principal axes of the inertial tensor of the spacecraft. As in Reference [15], we choose one relative equilibrium as shown by Fig. 2 without of loss of generality to stabilize:

$$\overline{\boldsymbol{R}}_e = \boldsymbol{\alpha}_e = [1, 0, 0]^T, \; \boldsymbol{\beta}_e = [0, 1, 0]^T, \; \boldsymbol{\gamma}_e = [0, 0, 1]^T, \tag{14}$$

$$\boldsymbol{\Pi}_e = [0, 0, \omega_T I_{zz}]^T, \; \boldsymbol{P}_e = [0, m\omega_T R_e, 0]^T, \tag{15}$$

$$\omega_T^2 R_e = \frac{\mu}{R_e^2} - \frac{3\mu}{2R_e^4}\left\{2\frac{I_{xx}}{m} - \frac{I_{yy}}{m} - \frac{I_{zz}}{m} + \tau_0 - 6\tau_2\right\}, \tag{16}$$

$$\mu_1 = -\frac{6\mu m \tau_2}{R_e^3}, \; \mu_2 = 0, \; \mu_3 = -\omega_T^2 I_{zz} - m\omega_T^2 R_e^2, \; \mu_4 = \mu_5 = \mu_6 = 0. \tag{17}$$

## STABILITY CONDITIONS BY ENERGY-CASIMIR METHOD

The energy-Casimir method, generalization of the Lagrange-Dirichlet criterion, is a powerful tool provided by geometric mechanics for determining the nonlinear stability of the relative equilibria in a non-canonical Hamiltonian system[20]. According to the energy-Casimir method, the condition of nonlinear stability of the relative equilibrium $z_e$ can be obtained through the eigenvalues of the projected Hessian matrix of the variational Lagrangian $F(z)$. The projected Hessian matrix is given by $P(z_e)\nabla^2 F(z_e)P(z_e)$, and the projection operator $P(z_e)$ is

$$P(z_e) = \mathbf{I}_{18\times 18} - K(z_e)\left(K(z_e)^T K(z_e)\right)^{-1} K(z_e)^T, \tag{18}$$

$$K(z_e) = \left[\nabla_z C_1(z)\big|_e \quad \cdots \quad \nabla_z C_i(z)\big|_e \quad \cdots \quad \nabla_z C_6(z)\big|_e\right] = \begin{bmatrix} 0 & 0 & 0 & 0 & 0 & 0 \\ \alpha_e & 0 & 0 & \beta_e & \gamma_e & 0 \\ 0 & \beta_e & 0 & \alpha_e & 0 & \gamma_e \\ 0 & 0 & \gamma_e & 0 & \alpha_e & \beta_e \\ 0 & 0 & 0 & 0 & 0 & 0 \\ 0 & 0 & 0 & 0 & 0 & 0 \end{bmatrix}. \tag{19}$$

The Hessian matrix $\nabla^2 F(z_e)$ is given by[15]:

$$\nabla^2 F(z_e) = \begin{bmatrix} \mathbf{I}^{-1} & 0 & 0 & -\omega_T \mathbf{I}_{3\times 3} & 0 & 0 \\ 0 & \dfrac{6\mu m \tau_2}{R_e^3}\left(\mathbf{I}_{3\times 3} - \alpha_e \alpha_e^T\right) & 0 & 0 & \left(\dfrac{\partial^2 V}{\partial \boldsymbol{\alpha} \partial \mathbf{R}}\bigg|_e\right)^T & 0 \\ 0 & 0 & \dfrac{6\mu m \tau_2}{R_e^3}\alpha_e \alpha_e^T & 0 & \left(\dfrac{\partial^2 V}{\partial \boldsymbol{\beta} \partial \mathbf{R}}\bigg|_e\right)^T & 0 \\ -\omega_T \mathbf{I}_{3\times 3} & 0 & 0 & \omega_T^2\left(I_{zz} + mR_e^2\right)\mathbf{I}_{3\times 3} - \dfrac{3\mu m \tau_0}{R_e^3}\alpha_e \alpha_e^T & \left(\dfrac{\partial^2 V}{\partial \boldsymbol{\gamma} \partial \mathbf{R}}\bigg|_e\right)^T + \omega_T^2 m R_e \hat{\boldsymbol{\beta}}_e & -\omega_T R_e \hat{\boldsymbol{\alpha}}_e \\ 0 & \dfrac{\partial^2 V}{\partial \boldsymbol{\alpha} \partial \mathbf{R}}\bigg|_e & \dfrac{\partial^2 V}{\partial \boldsymbol{\beta} \partial \mathbf{R}}\bigg|_e & \dfrac{\partial^2 V}{\partial \boldsymbol{\gamma} \partial \mathbf{R}}\bigg|_e - \omega_T^2 m R_e \hat{\boldsymbol{\beta}}_e & \dfrac{\partial^2 V}{\partial \mathbf{R}^2}\bigg|_e & \omega_T \hat{\boldsymbol{\gamma}}_e \\ 0 & 0 & 0 & \omega_T R_e \hat{\boldsymbol{\alpha}}_e & -\omega_T \hat{\boldsymbol{\gamma}}_e & \dfrac{1}{m}\mathbf{I}_{3\times 3} \end{bmatrix}. \tag{20}$$

The second-order partial derivates of the potential in Eq. (20) are given as follows:

$$\left.\frac{\partial^2 V}{\partial \boldsymbol{\alpha} \partial \boldsymbol{R}}\right|_e = -\frac{6\mu m \tau_2}{R_e^4}\left[\mathbf{I}_{3\times 3} - 4\boldsymbol{\alpha}_e\boldsymbol{\alpha}_e^T\right], \tag{21}$$

$$\left.\frac{\partial^2 V}{\partial \boldsymbol{\beta} \partial \boldsymbol{R}}\right|_e = \frac{6\mu m \tau_2}{R_e^4}\boldsymbol{\beta}_e\boldsymbol{\alpha}_e^T, \tag{22}$$

$$\left.\frac{\partial^2 V}{\partial \boldsymbol{\gamma} \partial \boldsymbol{R}}\right|_e = -\frac{3\mu m \tau_0}{R_e^4}\boldsymbol{\gamma}_e\boldsymbol{\alpha}_e^T, \tag{23}$$

$$\left.\frac{\partial^2 V}{\partial \boldsymbol{R}^2}\right|_e = \frac{\mu m}{R_e^3}\left(\mathbf{I}_{3\times 3} - 3\boldsymbol{\alpha}_e\boldsymbol{\alpha}_e^T\right) + \frac{3\mu}{2R_e^5}\left\{\left[15 I_{xx} - 5tr(\boldsymbol{I}) + 5\tau_0 m - 34\tau_2 m\right]\boldsymbol{\alpha}_e\boldsymbol{\alpha}_e^T + 4\tau_2 m \boldsymbol{\beta}_e\boldsymbol{\beta}_e^T \right.$$
$$\left. -2\tau_0 m \boldsymbol{\gamma}_e\boldsymbol{\gamma}_e^T + 2\boldsymbol{I} - \left[5I_{xx} - tr(\boldsymbol{I}) + \tau_0 m - 10\tau_2 m\right]\mathbf{I}_{3\times 3}\right\}. \tag{24}$$

The projected Hessian matrix of the variational Lagrangian $F(z)$ has the same number of zero eigenvalues as the linearly independent Casimir functions, which are associated with the complement space of $T\Sigma|_{z_e}$, the tangent space to the invariant manifold. The remaining eigenvalues are associated with the tangent space $T\Sigma|_{z_e}$. If they are all positive, the relative equilibrium $z_e$ is a constrained minimum on the invariant manifold $\Sigma$ and therefore nonlinear stable.

The eigenvalues are roots of the characteristic equation, which is given by

$$\det\left[s\mathbf{I}_{18\times 18} - \boldsymbol{P}(z_e)\nabla^2 F(z_e)\boldsymbol{P}(z_e)\right] = 0. \tag{25}$$

With the help of the symbolic calculation in *Matlab* and *Maple*, the characteristic equation Eq. (25) can be obtained with the following form:

$$\begin{aligned} s^6(I_{zz}s - 1)(mR_e^5 s^2 + A_1 s + A_0)(2mR_e^8 s^3 + B_2 s^2 + B_1 s + B_0)\\ (4I_{xx}R_e^8 s^3 + C_2 s^2 + C_1 s + C_0)(2mI_{yy} s^3 + D_2 s^2 + D_1 s + D_0) = 0. \end{aligned} \tag{26}$$

where $A_1$, $A_0$, $B_2$, $B_1$, $B_0$, $C_2$, $C_1$, $C_0$ $D_2$, $D_1$ and $D_0$ can be found in Reference [21].

In our problem there are six linearly independent Casimir functions, and the projected Hessian matrix have six zero eigenvalues associated with the six-dimensional complement space of the tangent space $T\Sigma|_{z_e}$. Nonlinear stability of the relative equilibrium requires that the remaining twelve eigenvalues are all positive. According to the theory of roots of the polynomial equation, that the remaining twelve eigenvalues in Eq. (26) are positive is equivalent to

$$\begin{aligned} A_1 < 0, A_0 > 0, B_2 < 0, B_1 > 0, B_0 < 0,\\ C_2 < 0, C_1 > 0, C_0 < 0, D_2 < 0, D_1 > 0, D_0 < 0. \end{aligned} \tag{27}$$

The inequality (27) is the condition of the nonlinear stability of the relative equilibria in the uncontrolled non-canonical Hamiltonian system. In the next section, this criterion can be used to verify the Lyapunov stability of the potential shaping feedback control law, since the non-canonical Hamiltonian structure of the system is preserved by the potential shaping.

**POTENTIAL SHAPING FEEDBACK CONTROL LAW**

The nonlinear stability of the natural relative equilibria cannot be always guaranteed by the spacecraft. Therefore, the control effort to stabilize the unstable relative equilibria is necessary to achieve the body-fixed orbit-attitude hovering. In this and next section, a non-canonical Hamiltonian structure-based stabilization feedback control law will be derived by using the potential shaping and energy dissipation.

**Potential Shaping and Feedback Control Law**

The potential of a Hamiltonian system can be modified by an artificial potential introduced by feedback control. In the potential shaping method, the feedback control is well designed so that the nonlinear/Lyapunov stability of the relative equilibrium can be guaranteed by the modified potential. Assume that the gravitational potential of the system given by Eq. (7) has been modified by the feedback control as follows:

$$V_C(\boldsymbol{R}, \boldsymbol{\alpha}, \boldsymbol{\beta}, \boldsymbol{\gamma}) = V(\boldsymbol{R}, \boldsymbol{\alpha}, \boldsymbol{\beta}, \boldsymbol{\gamma}) + \Delta V(\boldsymbol{R} - \boldsymbol{R}_e, \boldsymbol{\alpha} - \boldsymbol{\alpha}_e, \boldsymbol{\beta} - \boldsymbol{\beta}_e, \boldsymbol{\gamma} - \boldsymbol{\gamma}_e), \tag{28}$$

where $V(\boldsymbol{R}, \boldsymbol{\alpha}, \boldsymbol{\beta}, \boldsymbol{\gamma})$ is the potential of the uncontrolled system in Eq. (7); the second term $\Delta V(\boldsymbol{R} - \boldsymbol{R}_e, \boldsymbol{\alpha} - \boldsymbol{\alpha}_e, \boldsymbol{\beta} - \boldsymbol{\beta}_e, \boldsymbol{\gamma} - \boldsymbol{\gamma}_e)$ is the artificial potential introduced by the potential shaping; the attitude $\boldsymbol{\alpha}_e$, $\boldsymbol{\beta}_e$, $\boldsymbol{\gamma}_e$ and the position $\boldsymbol{R}_e$ at the relative equilibrium are given by Eqs. (14) and (16).

According to Eqs. (10) and (11), the potential shaping feedback control torque and force are:

$$\boldsymbol{T}_{C1}(\delta\boldsymbol{R}, \delta\boldsymbol{\alpha}, \delta\boldsymbol{\beta}, \delta\boldsymbol{\gamma}) = \boldsymbol{R} \times \frac{\partial \Delta V(\delta\boldsymbol{R}, \delta\boldsymbol{\alpha}, \delta\boldsymbol{\beta}, \delta\boldsymbol{\gamma})}{\partial \delta \boldsymbol{R}} + \boldsymbol{\alpha} \times \frac{\partial \Delta V(\delta\boldsymbol{R}, \delta\boldsymbol{\alpha}, \delta\boldsymbol{\beta}, \delta\boldsymbol{\gamma})}{\partial \delta \boldsymbol{\alpha}} \\ + \boldsymbol{\beta} \times \frac{\partial \Delta V(\delta\boldsymbol{R}, \delta\boldsymbol{\alpha}, \delta\boldsymbol{\beta}, \delta\boldsymbol{\gamma})}{\partial \delta \boldsymbol{\beta}} + \boldsymbol{\gamma} \times \frac{\partial \Delta V(\delta\boldsymbol{R}, \delta\boldsymbol{\alpha}, \delta\boldsymbol{\beta}, \delta\boldsymbol{\gamma})}{\partial \delta \boldsymbol{\gamma}}, \tag{29}$$

$$\boldsymbol{F}_{C1}(\delta\boldsymbol{R}, \delta\boldsymbol{\alpha}, \delta\boldsymbol{\beta}, \delta\boldsymbol{\gamma}) = -\frac{\partial \Delta V(\delta\boldsymbol{R}, \delta\boldsymbol{\alpha}, \delta\boldsymbol{\beta}, \delta\boldsymbol{\gamma})}{\partial \delta \boldsymbol{R}}, \tag{30}$$

respectively, where $\delta\boldsymbol{R} = \boldsymbol{R} - \boldsymbol{R}_e$, $\delta\boldsymbol{\alpha} = \boldsymbol{\alpha} - \boldsymbol{\alpha}_e$, $\delta\boldsymbol{\beta} = \boldsymbol{\beta} - \boldsymbol{\beta}_e$ and $\delta\boldsymbol{\gamma} = \boldsymbol{\gamma} - \boldsymbol{\gamma}_e$.

Notice that the feedback control torque and force should be zero at the relative equilibrium to keep the relative equilibrium unshifted. Then we have

$$\left.\frac{\partial \Delta V(\delta\boldsymbol{R}, \delta\boldsymbol{\alpha}, \delta\boldsymbol{\beta}, \delta\boldsymbol{\gamma})}{\partial \delta \boldsymbol{R}}\right|_{\delta\boldsymbol{R}=\delta\boldsymbol{\alpha}=\delta\boldsymbol{\beta}=\delta\boldsymbol{\gamma}=0} = \boldsymbol{0}, \tag{31}$$

$$\left.\frac{\partial \Delta V(\delta\boldsymbol{R}, \delta\boldsymbol{\alpha}, \delta\boldsymbol{\beta}, \delta\boldsymbol{\gamma})}{\partial \delta \boldsymbol{\alpha}}\right|_{\delta\boldsymbol{R}=\delta\boldsymbol{\alpha}=\delta\boldsymbol{\beta}=\delta\boldsymbol{\gamma}=0} = \boldsymbol{0}. \tag{32}$$

$$\left.\frac{\partial \Delta V(\delta\boldsymbol{R}, \delta\boldsymbol{\alpha}, \delta\boldsymbol{\beta}, \delta\boldsymbol{\gamma})}{\partial \delta \boldsymbol{\beta}}\right|_{\delta\boldsymbol{R}=\delta\boldsymbol{\alpha}=\delta\boldsymbol{\beta}=\delta\boldsymbol{\gamma}=0} = \boldsymbol{0}, \tag{33}$$

$$\left.\frac{\partial \Delta V(\delta\boldsymbol{R}, \delta\boldsymbol{\alpha}, \delta\boldsymbol{\beta}, \delta\boldsymbol{\gamma})}{\partial \delta \boldsymbol{\gamma}}\right|_{\delta\boldsymbol{R}=\delta\boldsymbol{\alpha}=\delta\boldsymbol{\beta}=\delta\boldsymbol{\gamma}=0} = \boldsymbol{0}. \tag{34}$$

According to Eqs. (6) and (28), the controlled Hamiltonian $H_C$ is given by:

$$H_C(\boldsymbol{z}) = \boldsymbol{\Pi}^T \boldsymbol{I}^{-1} \boldsymbol{\Pi}/2 + |\boldsymbol{P}|^2/2m - \boldsymbol{\omega}_T \boldsymbol{\Pi}^T \boldsymbol{\gamma} - \boldsymbol{\omega}_T \boldsymbol{P}^T (\hat{\boldsymbol{\gamma}} \boldsymbol{R}) + V_C(\boldsymbol{z}), \tag{35}$$

where the modified potential $V_C$ is given by Eq. (28). The controlled variational Lagrangian is:

$$F_C(\boldsymbol{z}) = H_C(\boldsymbol{z}) - \sum_{i=1}^{6} \mu_i C_i(\boldsymbol{z}) = F(\boldsymbol{z}) + \Delta V(\delta\boldsymbol{R}, \delta\boldsymbol{\alpha}, \delta\boldsymbol{\beta}, \delta\boldsymbol{\gamma}), \tag{36}$$

where $F(z)$ is the variational Lagrangian of the uncontrolled system. The non-canonical Hamiltonian structure of the system, including the Poisson tensor $B(z)$, Casimir functions $C_i(z)$ and the value of $\mu_i$, are preserved by the potential shaping. The Hessian matrix of the controlled variational Lagrangian $\nabla^2 F_C(z_e)$ at the relative equilibrium $z_e$ is given by:

$$\nabla^2 F_C(z_e) = \nabla^2 F(z_e) + \nabla^2 \Delta V(z_e), \tag{37}$$

where the Hessian matrix of the uncontrolled variational Lagrangian $\nabla^2 F(z_e)$ is given by Eq. (20). The Hessian matrix of the artificial potential $\nabla^2 \Delta V(z_e)$ at the relative equilibrium $z_e$ is:

$$\nabla^2 \Delta V(z_e) = \begin{bmatrix} 0 & 0 & 0 & 0 & 0 & 0 \\ 0 & \left.\frac{\partial^2 \Delta V(\delta z)}{\partial \delta \boldsymbol{\alpha}^2}\right|_{\delta z=0} & \left(\left.\frac{\partial^2 \Delta V(\delta z)}{\partial \delta \boldsymbol{\alpha} \delta \boldsymbol{\beta}}\right|_{\delta z=0}\right)^T & \left(\left.\frac{\partial^2 \Delta V(\delta z)}{\partial \delta \boldsymbol{\alpha} \delta \boldsymbol{\gamma}}\right|_{\delta z=0}\right)^T & \left(\left.\frac{\partial^2 \Delta V(\delta z)}{\partial \delta \boldsymbol{\alpha} \delta \boldsymbol{R}}\right|_{\delta z=0}\right)^T & 0 \\ 0 & \left.\frac{\partial^2 \Delta V(\delta z)}{\partial \delta \boldsymbol{\alpha} \delta \boldsymbol{\beta}}\right|_{\delta z=0} & \left.\frac{\partial^2 \Delta V(\delta z)}{\partial \delta \boldsymbol{\beta}^2}\right|_{\delta z=0} & \left(\left.\frac{\partial^2 \Delta V(\delta z)}{\partial \delta \boldsymbol{\beta} \delta \boldsymbol{\gamma}}\right|_{\delta z=0}\right)^T & \left(\left.\frac{\partial^2 \Delta V(\delta z)}{\partial \delta \boldsymbol{\beta} \delta \boldsymbol{R}}\right|_{\delta z=0}\right)^T & 0 \\ 0 & \left.\frac{\partial^2 \Delta V(\delta z)}{\partial \delta \boldsymbol{\alpha} \delta \boldsymbol{\gamma}}\right|_{\delta z=0} & \left.\frac{\partial^2 \Delta V(\delta z)}{\partial \delta \boldsymbol{\beta} \delta \boldsymbol{\gamma}}\right|_{\delta z=0} & \left.\frac{\partial^2 \Delta V(\delta z)}{\partial \delta \boldsymbol{\gamma}^2}\right|_{\delta z=0} & \left(\left.\frac{\partial^2 \Delta V(\delta z)}{\partial \delta \boldsymbol{\gamma} \delta \boldsymbol{R}}\right|_{\delta z=0}\right)^T & 0 \\ 0 & \left.\frac{\partial^2 \Delta V(\delta z)}{\partial \delta \boldsymbol{\alpha} \delta \boldsymbol{R}}\right|_{\delta z=0} & \left.\frac{\partial^2 \Delta V(\delta z)}{\partial \delta \boldsymbol{\beta} \delta \boldsymbol{R}}\right|_{\delta z=0} & \left.\frac{\partial^2 \Delta V(\delta z)}{\partial \delta \boldsymbol{\gamma} \delta \boldsymbol{R}}\right|_{\delta z=0} & \left.\frac{\partial^2 \Delta V(\delta z)}{\partial \delta \boldsymbol{R}^2}\right|_{\delta z=0} & 0 \\ 0 & 0 & 0 & 0 & 0 & 0 \end{bmatrix}. \tag{38}$$

Notice that the second-order partial derivates of the terms higher than second order in the artificial potential $\Delta V(\delta z)$ are all equal to zero at the relative equilibrium. According to Eq. (38), the terms higher than second order in the artificial potential $\Delta V(\delta z)$ have no effect on the stability of the relative equilibrium. From Eqs. (31)-(34), we can know that there is no first-order terms in the artificial potential $\Delta V(\delta z)$. Therefore, the artificial potential $\Delta V(\delta z)$ can be chosen as a quadratic form of the attitude error $\delta \boldsymbol{\alpha}$, $\delta \boldsymbol{\beta}$, $\delta \boldsymbol{\gamma}$ and the position error $\delta \boldsymbol{R}$, which can be written as:

$$\Delta V(\delta \boldsymbol{R}, \delta \boldsymbol{\alpha}, \delta \boldsymbol{\beta}, \delta \boldsymbol{\gamma}) = \frac{1}{2}\begin{bmatrix} \delta \boldsymbol{\alpha}^T, \delta \boldsymbol{\beta}^T, \delta \boldsymbol{\gamma}^T, \delta \boldsymbol{R}^T \end{bmatrix} \begin{bmatrix} W_{\alpha\alpha} & W_{\alpha\beta}^T & W_{\alpha\gamma}^T & W_{\alpha R}^T \\ W_{\alpha\beta} & W_{\beta\beta} & W_{\beta\gamma}^T & W_{\beta R}^T \\ W_{\alpha\gamma} & W_{\beta\gamma} & W_{\gamma\gamma} & W_{\gamma R}^T \\ W_{\alpha R} & W_{\beta R} & W_{\gamma R} & W_{RR} \end{bmatrix} \begin{bmatrix} \delta \boldsymbol{\alpha} \\ \delta \boldsymbol{\beta} \\ \delta \boldsymbol{\gamma} \\ \delta \boldsymbol{R} \end{bmatrix}, \tag{39}$$

where the matrices $W_{\alpha\alpha}$, $W_{\beta\beta}$, $W_{\gamma\gamma}$, and $W_{RR}$ are all symmetric. Then $\nabla^2 \Delta V(z_e)$ in Eq. (38) is:

$$\nabla^2 \Delta V(z_e) = \begin{bmatrix} 0 & 0 & 0 & 0 & 0 & 0 \\ 0 & W_{\alpha\alpha} & W_{\alpha\beta}^T & W_{\alpha\gamma}^T & W_{\alpha R}^T & 0 \\ 0 & W_{\alpha\beta} & W_{\beta\beta} & W_{\beta\gamma}^T & W_{\beta R}^T & 0 \\ 0 & W_{\alpha\gamma} & W_{\beta\gamma} & W_{\gamma\gamma} & W_{\gamma R}^T & 0 \\ 0 & W_{\alpha R} & W_{\beta R} & W_{\gamma R} & W_{RR} & 0 \\ 0 & 0 & 0 & 0 & 0 & 0 \end{bmatrix}. \tag{40}$$

According to Eqs. (29) and (30), the potential shaping feedback control law are given by:

$$T_{C1}(\delta \boldsymbol{R}, \delta \boldsymbol{\alpha}, \delta \boldsymbol{\beta}, \delta \boldsymbol{\gamma}) =$$
$$(\boldsymbol{R}_e + \delta \boldsymbol{R}) \times (W_{\alpha R} \delta \boldsymbol{\alpha} + W_{\beta R} \delta \boldsymbol{\beta} + W_{\gamma R} \delta \boldsymbol{\gamma} + W_{RR} \delta \boldsymbol{R}) + (\boldsymbol{\alpha}_e + \delta \boldsymbol{\alpha}) \times (W_{\alpha\alpha} \delta \boldsymbol{\alpha} + W_{\alpha\beta}^T \delta \boldsymbol{\beta} + W_{\alpha\gamma}^T \delta \boldsymbol{\gamma} + W_{\alpha R}^T \delta \boldsymbol{R}) \quad (41)$$
$$+ (\boldsymbol{\beta}_e + \delta \boldsymbol{\beta}) \times (W_{\alpha\beta} \delta \boldsymbol{\alpha} + W_{\beta\beta} \delta \boldsymbol{\beta} + W_{\beta\gamma}^T \delta \boldsymbol{\gamma} + W_{\beta R}^T \delta \boldsymbol{R}) + (\boldsymbol{\gamma}_e + \delta \boldsymbol{\gamma}) \times (W_{\alpha\gamma} \delta \boldsymbol{\alpha} + W_{\beta\gamma} \delta \boldsymbol{\beta} + W_{\gamma\gamma} \delta \boldsymbol{\gamma} + W_{\gamma R}^T \delta \boldsymbol{R}),$$

$$F_{C1}(\delta \boldsymbol{R}, \delta \boldsymbol{\alpha}, \delta \boldsymbol{\beta}, \delta \boldsymbol{\gamma}) = -(W_{\alpha R} \delta \boldsymbol{\alpha} + W_{\beta R} \delta \boldsymbol{\beta} + W_{\gamma R} \delta \boldsymbol{\gamma} + W_{RR} \delta \boldsymbol{R}). \quad (42)$$

**Lyapunov Stability**

The Lyapunov stability of the system with potential shaping feedback control law Eqs. (41) and (42) can be verified by using the energy-Casimir method, since the non-canonical Hamiltonian structure of the system is preserved by the potential shaping, as shown before. The Lyapunov stability of the potential shaping feedback control law is equivalent to the nonlinear stability of the non-canonical Hamiltonian system with the modified potential $V_C(\boldsymbol{R}, \boldsymbol{\alpha}, \boldsymbol{\beta}, \boldsymbol{\gamma})$.

When the potential shaping feedback control law Eqs. (41) and (42) is acted on the system, the artificial potential $\Delta V(\delta \boldsymbol{R}, \delta \boldsymbol{\alpha}, \delta \boldsymbol{\beta}, \delta \boldsymbol{\gamma})$ given by Eq. (39) is added into the potential of the original system equivalently. According to Eqs. (20), (37), and (40), the Hessian matrix of the controlled variational Lagrangian $\nabla^2 F_C(z_e)$ at the relative equilibrium $z_e$ is given by:

$$\nabla^2 F_C(z_e) = \begin{bmatrix} \boldsymbol{I}^{-1} & 0 & 0 \\ 0 & \frac{6\mu m \tau_2}{R_e^3}(\boldsymbol{I}_{3\times 3} - \boldsymbol{\alpha}_e \boldsymbol{\alpha}_e^T) + W_{\alpha\alpha} & W_{\alpha\beta}^T \\ 0 & W_{\alpha\beta} & \frac{6\mu m \tau_2}{R_e^3} \boldsymbol{\alpha}_e \boldsymbol{\alpha}_e^T + W_{\beta\beta} \\ -\omega_T \boldsymbol{I}_{3\times 3} & W_{\alpha\gamma} & W_{\beta\gamma} \\ 0 & \left.\frac{\partial^2 V}{\partial \boldsymbol{\alpha} \partial \boldsymbol{R}}\right|_e + W_{\alpha R} & \left.\frac{\partial^2 V}{\partial \boldsymbol{\beta} \partial \boldsymbol{R}}\right|_e + W_{\beta R} \\ 0 & 0 & 0 \end{bmatrix}$$

$$\begin{matrix} -\omega_T \boldsymbol{I}_{3\times 3} & 0 & 0 \\ W_{\alpha\gamma}^T & \left(\left.\frac{\partial^2 V}{\partial \boldsymbol{\alpha} \partial \boldsymbol{R}}\right|_e\right)^T + W_{\alpha R}^T & 0 \\ W_{\beta\gamma}^T & \left(\left.\frac{\partial^2 V}{\partial \boldsymbol{\beta} \partial \boldsymbol{R}}\right|_e\right)^T + W_{\beta R}^T & 0 \\ \omega_T^2 (I_{zz} + m R_e^2) \boldsymbol{I}_{3\times 3} - \frac{3\mu m \tau_0}{R_e^3} \boldsymbol{\alpha}_e \boldsymbol{\alpha}_e^T + W_{\gamma\gamma} & \left(\left.\frac{\partial^2 V}{\partial \boldsymbol{\gamma} \partial \boldsymbol{R}}\right|_e\right)^T + \omega_T^2 m R_e \hat{\boldsymbol{\beta}}_e + W_{\gamma R}^T & -\omega_T R_e \hat{\boldsymbol{\alpha}}_e \\ \left.\frac{\partial^2 V}{\partial \boldsymbol{\gamma} \partial \boldsymbol{R}}\right|_e - \omega_T^2 m R_e \hat{\boldsymbol{\beta}}_e + W_{\gamma R} & \left.\frac{\partial^2 V}{\partial \boldsymbol{R}^2}\right|_e + W_{RR} & \omega_T \hat{\boldsymbol{\gamma}}_e \\ \omega_T R_e \hat{\boldsymbol{\alpha}}_e & -\omega_T \hat{\boldsymbol{\gamma}}_e & \frac{1}{m} \boldsymbol{I}_{3\times 3} \end{matrix}. \quad (43)$$

The projected Hessian matrix of the controlled variational Lagrangian is given by $\boldsymbol{P}(z_e) \nabla^2 F_C(z_e) \boldsymbol{P}(z_e)$. The projection operator $\boldsymbol{P}(z_e)$ is still given by Eq. (18), since the potential shaping preserves the Casimir functions $C_i(z)$. According to the energy-Casimir method, the Lyapunov stability of the potential shaping feedback control law implies that eigenvalues of the projected Hessian matrix of controlled variational Lagrangian $P(z_e)\nabla^2 F_C(z_e)P(z_e)$ are six zeros and twelve positive real numbers. The eigenvalues are roots of the characteristic equation

$$\det\left[s\mathbf{I}_{18\times 18} - P(z_e)\nabla^2 F_C(z_e)P(z_e)\right] = 0. \tag{44}$$

To simplify the control design, here we adopt a very simple form for the artificial potential $\Delta V(\delta R, \delta\alpha, \delta\beta, \delta\gamma)$, which is given by:

$$\Delta V(\delta R, \delta\alpha, \delta\beta, \delta\gamma) = \frac{1}{2}\left[\delta\alpha^T, \delta\beta^T, \delta\gamma^T, \delta R^T\right]\begin{bmatrix} W_{\alpha\alpha} & 0 & 0 & 0 \\ 0 & W_{\beta\beta} & 0 & 0 \\ 0 & 0 & W_{\gamma\gamma} & 0 \\ 0 & 0 & 0 & W_{RR} \end{bmatrix}\begin{bmatrix} \delta\alpha \\ \delta\beta \\ \delta\gamma \\ \delta R \end{bmatrix}, \tag{45}$$

$$W_{\alpha\alpha} = \begin{bmatrix} \eta_1 & 0 & 0 \\ 0 & \eta_2 & 0 \\ 0 & 0 & \eta_3 \end{bmatrix}, \quad W_{\beta\beta} = \begin{bmatrix} \eta_4 & 0 & 0 \\ 0 & \eta_5 & 0 \\ 0 & 0 & \eta_6 \end{bmatrix}, \quad W_{\gamma\gamma} = \begin{bmatrix} \eta_7 & 0 & 0 \\ 0 & \eta_8 & 0 \\ 0 & 0 & \eta_9 \end{bmatrix}, \tag{46}$$

$$W_{RR} = \begin{bmatrix} \eta_{10} & 0 & 0 \\ 0 & \eta_{11} & 0 \\ 0 & 0 & \eta_{12} \end{bmatrix}. \tag{47}$$

Through some calculation, we find that among the feedback gains only the terms $\eta_2 + \eta_4$, $\eta_3 + \eta_7$, $\eta_6 + \eta_8$, $\eta_{10}$, $\eta_{11}$ and $\eta_{12}$ appear in the projected Hessian matrix $P(z_e)\nabla^2 F_C(z_e)P(z_e)$. Therefore, the feedback gains $W_{\alpha\alpha}$, $W_{\beta\beta}$ and $W_{\gamma\gamma}$ in Eq. (46) can be chosen as:

$$W_{\alpha\alpha} = \begin{bmatrix} 0 & 0 & 0 \\ 0 & \eta_2 & 0 \\ 0 & 0 & \eta_3 \end{bmatrix}, \quad W_{\beta\beta} = \begin{bmatrix} 0 & 0 & 0 \\ 0 & 0 & 0 \\ 0 & 0 & \eta_6 \end{bmatrix}, \quad W_{\gamma\gamma} = \begin{bmatrix} 0 & 0 & 0 \\ 0 & 0 & 0 \\ 0 & 0 & 0 \end{bmatrix}. \tag{48}$$

That is to say, only three feedback gains $\eta_2$, $\eta_3$ and $\eta_6$ are adopted in the attitude feedback that is reasonable since there are only three independent degrees of freedom in the attitude motion parameters $\alpha$, $\beta$ and $\gamma$. The characteristic equation of the projected Hessian matrix Eq. (44) with the artificial potential Eqs. (45), (47) and (48) is obtained with the following form:

$$s^6(I_{zz}s - 1)(mR_e^5 s^2 + A_{1C}s + A_{0C})(2mR_e^8 s^3 + B_{2C}s^2 + B_{1C}s + B_{0C}) \\ (4I_{xx}R_e^8 s^3 + C_{2C}s^2 + C_{1C}s + C_{0C})(2mI_{yy}s^3 + D_{2C}s^2 + D_{1C}s + D_{0C}) = 0. \tag{49}$$

Through the comparison between the coefficients of the uncontrolled system in Eq. (26) and controlled system in Eq. (49), we find that the coefficients in Eq. (49) can be written as follows:

$$A_{1C} = A_1 - \eta_{10}mR_e^5, \tag{50}$$

$$A_{0C} = A_0 + R_e^5\eta_{10}, \tag{51}$$

$$B_{2C} = B_2 - 2\eta_{11}R_e^8 m - \eta_2 R_e^8 m, \tag{52}$$

$$B_{1C} = B_1 + \eta_{11}\eta_2 R_e^8 m + 2\eta_{11}R_e^5\left(R_e^3 + 6m^2\mu\tau_2\right) \\ + \eta_2 R_e^3\left(m^2 R_e^2\mu + 21m^2\mu\tau_2 - 1.5m^2\mu\tau_0 + R_e^5 + 4.5m\mu I_{yy} - 6mI_{xx}\mu + 1.5m\mu I_{zz}\right), \tag{53}$$

$$B_{0C} = B_0 - \eta_{11}\eta_2 R_e^8 - 12\mu m R_e^5 \tau_2 \eta_{11}$$
$$+\eta_2 R_e^3 \left(6I_{xx}\mu - mR_e^2\mu - 21m\mu\tau_2 + 1.5m\mu\tau_0 + m\omega_T^2 R_e^5 - 4.5\mu I_{yy} - 1.5\mu I_{zz}\right), \tag{54}$$

$$C_{2C} = C_2 - 2\eta_3 I_{xx} R_e^8 - 4\eta_{12} I_{xx} R_e^8, \tag{55}$$

$$C_{1C} = C_1 + 2\eta_{12}\eta_3 I_{xx} R_e^8 + \eta_3 R_e^3 \left(30mI_{xx}\mu\tau_2 + 9I_{xx}\mu I_{zz} - 9mI_{xx}\mu\tau_0 - 12I_{xx}^2\mu + 2mR_e^2 I_{xx}\mu + 3I_{xx}\mu I_{yy} + 2R_e^5\right)$$
$$+\eta_{12} R_e^5 \left(4R_e^3 + 2m\omega_T^2 I_{xx} R_e^5 + 2R_e^3 \omega_T^2 I_{xx} I_{zz} - 6mI_{xx}\mu\tau_0 + 12mI_{xx}\mu\tau_2\right), \tag{56}$$

$$C_{0C} = C_0 - 2\eta_{12}\eta_3 R_e^8 + \eta_{12} R_e^5 \left(6m\mu\tau_0 - 12m\mu\tau_2 - 2m\omega_T^2 R_e^5 - 2R_e^3 \omega_T^2 I_{zz} + 2R_e^3 \omega_T^2 I_{xx}\right)$$
$$+\eta_3 R_e^3 \left(12I_{xx}\mu - 3\mu I_{yy} - 2mR_e^2\mu - 9\mu I_{zz} + 9m\mu\tau_0 - 30m\mu\tau_2\right), \tag{57}$$

$$D_{2C} = D_2 - \eta_6 m I_{yy}, \tag{58}$$

$$D_{1C} = D_1 + \eta_6 m + \eta_6 I_{yy}, \tag{59}$$

$$D_{0C} = D_0 - \eta_6. \tag{60}$$

According to Eqs. (26) and (49), we can see that the form of the characteristic equation of the projected Hessian matrix is preserved by the potential shaping. The condition of the Lyapunov stability of the potential shaping feedback control law is same with that of the nonlinear stability of the uncontrolled system Eq. (27):

$$A_{1C} < 0,\ A_{0C} > 0,\ B_{2C} < 0,\ B_{1C} > 0,\ B_{0C} < 0,$$
$$C_{2C} < 0,\ C_{1C} > 0,\ C_{0C} < 0,\ D_{2C} < 0,\ D_{1C} > 0,\ D_{0C} < 0. \tag{61}$$

According to Eqs. (50)-(60) and the condition of the Lyapunov stability inequality (61), the coefficients $A_{1C}$, $A_{0C}$, $B_{2C}$, $B_{1C}$, $B_{0C}$, $C_{2C}$, $C_{1C}$, $C_{0C}$ $D_{2C}$, $D_{1C}$ and $D_{0C}$ can satisfy the condition of the Lyapunov stability if the feedback gains $\eta_2$, $\eta_3$, $\eta_6$, $\eta_{10}$, $\eta_{11}$ and $\eta_{12}$ are chosen sufficiently large. That is to say, the unstable relative equilibrium can always be stabilized in the Lyapunov sense by potential shaping with sufficiently large feedback gains.

**Numerical Simulation**

The feasibility of the proposed potential shaping feedback control law will be verified through a numerical simulation in the case of a spacecraft orbiting around a small asteroid. The parameters of the asteroid and the spacecraft are chosen as follows:

$$\mu = 94\text{m}^3/\text{s}^2,\ C_{20} = -0.2,\ C_{22} = 0.2,\ a_e = 500\text{m}, \tag{62}$$

$$\tau_0 = -50000\text{m}^2,\ \tau_2 = 50000\text{m}^2,\ \omega_T = 2.9089\times 10^{-4}\text{s}^{-1}. \tag{63}$$

$$m = 5\times 10^3\text{kg},\ I_{xx}/m = 50\text{m}^2,\ \sigma_x = 0.3,\ \sigma_y = -0.4, \tag{64}$$

where the mass distribution parameters $\sigma_x$ and $\sigma_y$ are defined as

$$\sigma_x = \left(\frac{I_{zz} - I_{yy}}{I_{xx}}\right),\ \sigma_y = \left(\frac{I_{zz} - I_{xx}}{I_{yy}}\right). \tag{65}$$

The orbital radius of the relative equilibrium $R_e$ can be calculated by Eq. (16): $R_e = 1156.5$m. Then the relative equilibrium $z_e$ can be given by Eqs. (14)-(17). The coefficients $A_1$, $A_0$, $B_2$, $B_1$, $B_0$, $C_2$, $C_1$, $C_0$ $D_2$, $D_1$ and $D_0$ in Eq. (26) can by calculated as follows:

$A_1 = 9.1519 \times 10^{15}$, $A_0 = -3.1195 \times 10^{12}$, $B_2 = -2.9173 \times 10^{30}$, $B_1 = 1.8177 \times 10^{27}$, $B_0 = 9.9405 \times 10^{18}$, $C_2 = -1.1244 \times 10^{33}$, $C_1 = 4.504 \times 10^{27}$, $C_0 = 4.4876 \times 10^{19}$, $D_2 = -3.5369 \times 10^{11}$, $D_1 = 2.8316 \times 10^{6}$, $D_0 = -0.0063$.

Coefficients $A_1$, $A_0$, $B_0$ and $C_0$ do not satisfy the condition of the nonlinear stability Eq. (27), then the relative equilibrium $z_e$ is unstable. We implement our proposed potential shaping feedback control law to stabilize this unstable relative equilibrium. The feedback gains are chosen as:

$$\eta_2 = 1 \times 10^{-4}, \ \eta_3 = 1 \times 10^{-4}, \ \eta_6 = 1 \times 10^{-4}, \ \eta_{10} = 0.01, \ \eta_{11} = 1 \times 10^{-4}, \ \eta_{12} = 1 \times 10^{-4}. \tag{66}$$

The coefficients $A_{1C}$, $A_{0C}$, $B_{2C}$, $B_{1C}$, $B_{0C}$, $C_{2C}$, $C_{1C}$, $C_{0C}$, $D_{2C}$, $D_{1C}$ and $D_{0C}$ in the characteristic equation of the controlled system in Eq. (49) can by calculated as:

$A_{1C} = -9.4299 \times 10^{16}$, $A_{0C} = 1.7571 \times 10^{13}$, $B_{2C} = -2.9173 \times 10^{30}$, $B_{1C} = 2.1094 \times 10^{27}$, $B_{0C} = -5.8336 \times 10^{22}$, $C_{2C} = -1.1244 \times 10^{33}$, $C_{1C} = 1.1694 \times 10^{29}$, $C_{0C} = -4.4970 \times 10^{23}$, $D_{2C} = -3.5369 \times 10^{11}$, $D_{1C} = 2.8316 \times 10^{6}$, $D_{0C} = -0.0064$.

These coefficients satisfy the Lyapunov stability Eq. (61). Initial conditions of the motion are:

$$\boldsymbol{\Pi}_0 = \Omega_e I_{zz} [0, 0, 1]^T, \ \boldsymbol{\alpha}_0 = [\cos(\pi/18), \sin(\pi/18), 0]^T, \ \boldsymbol{\beta}_0 = [-\sin(\pi/18), \cos(\pi/18), 0]^T,$$
$$\boldsymbol{\gamma}_0 = [0, 0, 1]^T, \ \boldsymbol{R}_0 = R_e [1, 0.03, 0.03]^T, \ \boldsymbol{P}_0 = m R_e \Omega_e [0, 1, 0]^T. \tag{67}$$

The motion of the uncontrolled system, including $\boldsymbol{\Pi}$, $\boldsymbol{\alpha}$, $\boldsymbol{\beta}$, $\boldsymbol{\gamma}$, $\boldsymbol{P}$ and $\boldsymbol{R}$, are shown by Figs. 3-8 respectively. We can see that relative equilibrium $z_e$ of the uncontrolled system is unstable.

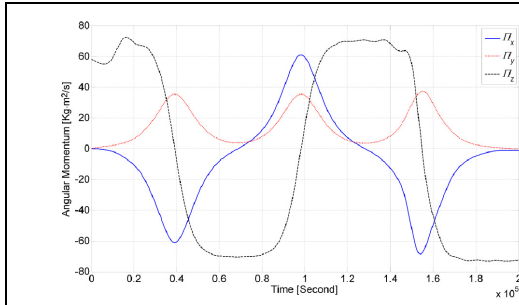

**Fig. 3. Angular momentum $\boldsymbol{\Pi}$ without control.**

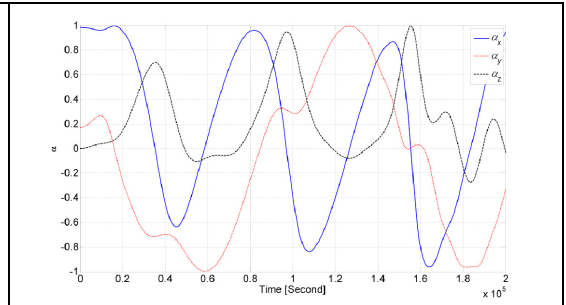

**Fig. 4. Attitude parameter $\boldsymbol{\alpha}$ without control.**

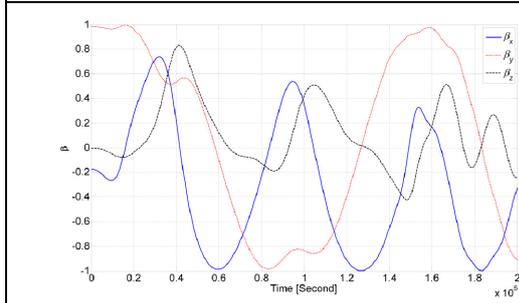

**Fig. 5. Attitude parameter $\boldsymbol{\beta}$ without control.**

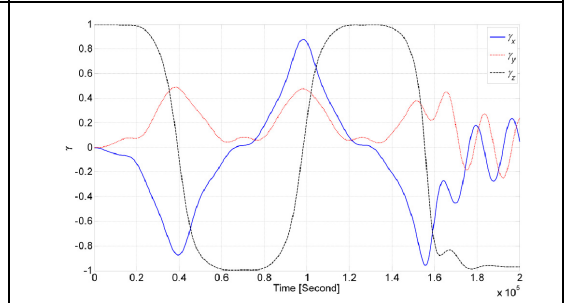

**Fig. 6. Attitude parameter $\boldsymbol{\gamma}$ without control.**

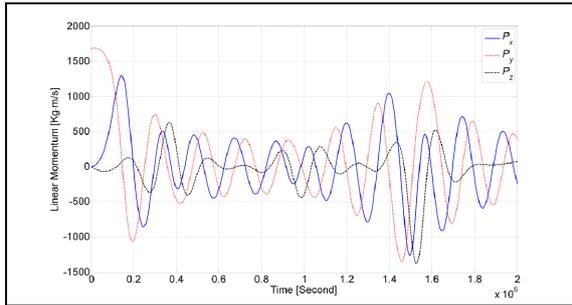

**Fig. 7. Linear momentum $P$ without control.**

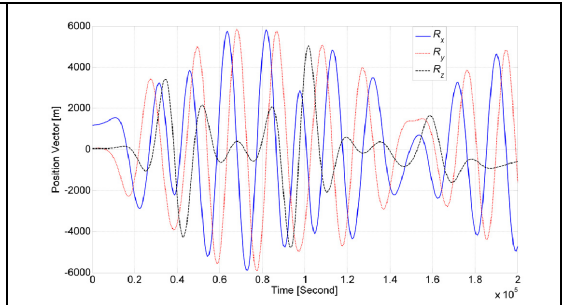

**Fig. 8. Position vector $R$ without control.**

After the introduction of the potential shaping feedback control law given by Eq. (66), the motion of the spacecraft, including $\Pi$, $\alpha$, $\beta$, $\gamma$, $P$ and $R$, are shown by Figs. 9-14 respectively. According to Figs. 9-14, it is easy to find that the relative equilibrium is successfully stabilized by our proposed potential shaping feedback control law, then the feasibility of which has been verified. However, since the system with the potential shaping feedback is still a Hamiltonian system, only the Lyapunov stability can be achieved by the potential shaping, but the asymptotic stability cannot be achieved, as shown by Figs. 9-14.

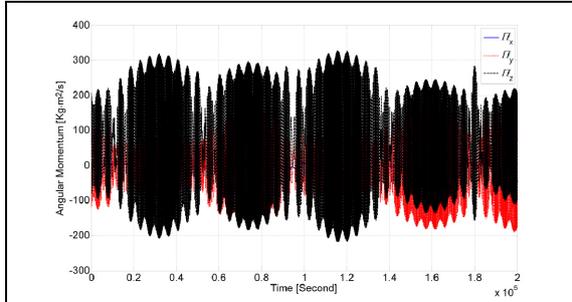

**Fig. 9. $\Pi$ with potential shaping.**

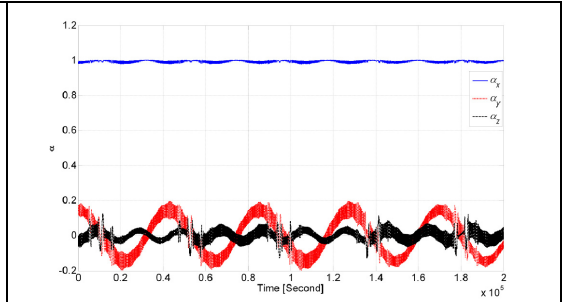

**Fig. 10. $\alpha$ with potential shaping.**

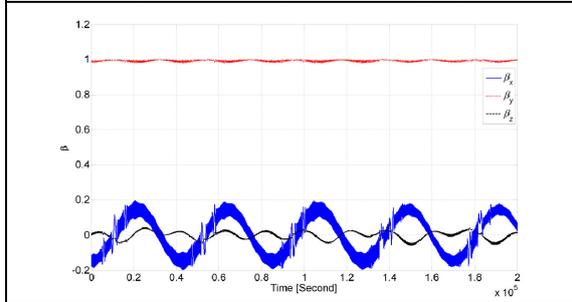

**Fig. 11. $\beta$ with potential shaping.**

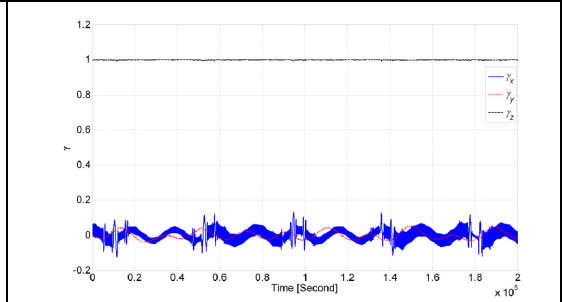

**Fig. 12. $\gamma$ with potential shaping.**

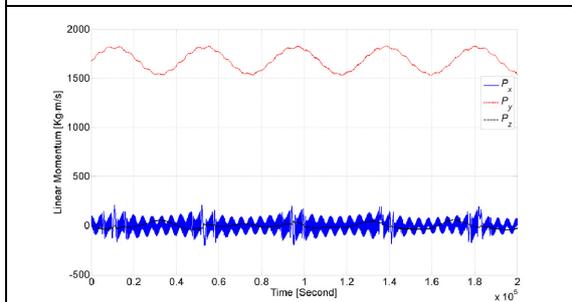

**Fig. 13. $P$ with potential shaping.**

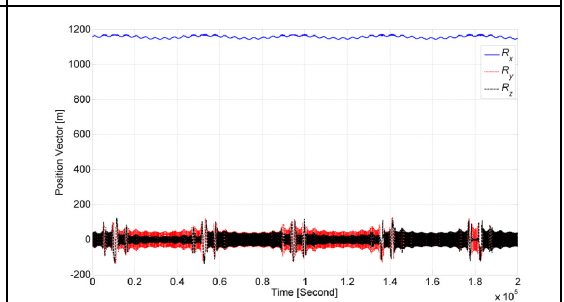

**Fig. 14. $R$ with potential shaping.**

The potential shaping control torque and force are shown by Figs. 15 and 16 respectively. The control torque and force cannot converge to the zero, since the motion of the system cannot be stabilized asymptotically by the potential shaping.

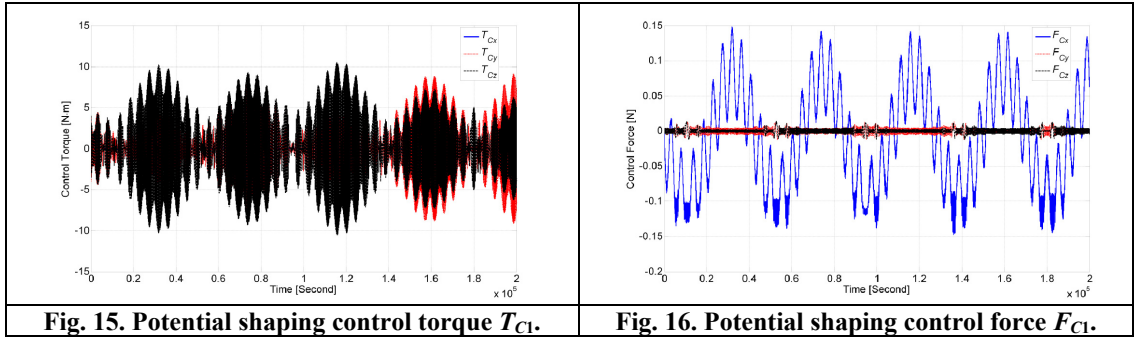

**Fig. 15. Potential shaping control torque $T_{C1}$.**     **Fig. 16. Potential shaping control force $F_{C1}$.**

## FEEDBACK CONTROL LAW WITH ENERGY DISSIPATION

Although the relative equilibrium $z_e$ has been successfully stabilized by our proposed potential shaping feedback control law, only the Lyapunov stability, but not the asymptotic stability, has been achieved. In this section, we will introduce the energy dissipation into the feedback control law to stabilize the relative equilibrium asymptotically.

### Energy Dissipation

From the geometrical viewpoint, Lyapunov stability of the potential shaping feedback control law, i.e., nonlinear stability of the system with modified potential, means that the modified energy is a minimum on the invariant manifold at the relative equilibrium[18]. The motion of the potential-shaped system is always staying on the twelve-dimensional invariant manifold $\Sigma$ given by Eq. (5). Since all the Casimir functions in Eq. (5) are geometric integrals, the phase flow of the system is constrained on the same invariant manifold with the relative equilibrium naturally by the equations of motion. The energy of the system $F_C(z)$ is larger than the minimum energy at $z_e$, therefore the system with the potential shaping will always librate around, but cannot converge to the relative equilibrium $z_e$, as shown by Figs. 9-14.

If the energy $F_C(z)$ can be dissipated by the control effort besides the potential shaping, the motion of the system will evolve to approaching the minimum energy point on the invariant manifold $\Sigma$, i.e., the relative equilibrium. Then the asymptotic stabilization of the relative equilibrium can be achieved. This is the basic principle of the Hamiltonian structure-based stabilization approach. The energy dissipation control torque $T_{C2}(z)$ and force $F_{C2}(z)$ can be designed as:

$$\boldsymbol{T}_{C2}(z) = \eta_{13}\left[F_C(z) - F_C(z_e)\right]\frac{\boldsymbol{I}^{-1}\boldsymbol{\Pi} - \omega_T\boldsymbol{\gamma}}{\left|\boldsymbol{I}^{-1}\boldsymbol{\Pi} - \omega_T\boldsymbol{\gamma}\right|}, \tag{68}$$

$$\boldsymbol{F}_{C2}(z) = \eta_{13}\left[F_C(z) - F_C(z_e)\right]\frac{\boldsymbol{P}/m - \omega_T\boldsymbol{\gamma}\times\boldsymbol{R}}{\left|\boldsymbol{P}/m - \omega_T\boldsymbol{\gamma}\times\boldsymbol{R}\right|}. \tag{69}$$

According to Eq. (6), the energy dissipation introduced by $\boldsymbol{T}_{C2}(z)$ and $\boldsymbol{F}_{C2}(z)$ can be given by

$$\begin{aligned}\frac{d}{dt}\left(F_C(z)\right) &= \left|\boldsymbol{T}_{C2}(z)\cdot\left(\boldsymbol{I}^{-1}\boldsymbol{\Pi} - \omega_T\boldsymbol{\gamma}\right)\right| + \left|\boldsymbol{F}_{C2}(z)\cdot\left(\boldsymbol{P}/m - \omega_T\boldsymbol{\gamma}\times\boldsymbol{R}\right)\right| \\ &= \eta_{13}\left[F_C(z) - F_C(z_e)\right]\left(\left|\boldsymbol{I}^{-1}\boldsymbol{\Pi} - \omega_T\boldsymbol{\gamma}\right| + \left|\boldsymbol{P}/m - \omega_T\boldsymbol{\gamma}\times\boldsymbol{R}\right|\right).\end{aligned} \tag{70}$$

Therefore, the energy of the system $F_C(z)$ will converge to the value at the relative equilibrium $F_C(z_e)$, the minimum on the invariant manifold $\Sigma$, under the energy dissipation control $T_{C2}(z)$ and $F_{C2}(z)$. That is to say, the motion of the system will converge asymptotically to the relative equilibrium $z_e$ under the energy dissipation control efforts. As the motion of the system is converging to the relative equilibrium, both the terms $|I^{-1}\Pi - \omega_T\gamma|$ and $|P/m - \omega_T\gamma \times R|$ converge to zero. Therefore, the feedback gain $\eta_{13}$ needs to be adjusted to speed up the convergence as the motion converges to the relative equilibrium.

The resultant full Hamiltonian structure-based feedback control law is given by:

$$T_C(z) = T_{C1}(z) + T_{C2}(z), \quad (71)$$

$$F_C(z) = F_{C1}(z) + F_{C2}(z), \quad (72)$$

where $T_{C1}(z)$, $F_{C1}(z)$, $T_{C2}(z)$ and $F_{C2}(z)$ are given by Eqs. (41), (42), (68) and (69) respectively.

**Numerical Simulation**

To verify the feasibility of our Hamiltonian structure-based feedback control law Eqs. (71) and (72), a numerical simulation with the same parameters and initial conditions as in the last section is carried out. The feedback gain $\eta_{13}$ in energy dissipation control Eqs. (68) and (69) is chosen to be $-1.6^{t/1\times 10^4} \times 10^{-2}$, where the factor $1.6^{t/1\times 10^4}$ is used to speed up the convergence when near to the relative equilibrium. The motion of the spacecraft, including $\Pi$, $\alpha$, $\beta$, $\gamma$, $P$ and $R$, under the full feedback control with potential shaping and energy dissipation are shown by Figs. 17-22 respectively. The simulation results show that the relative equilibrium has been successfully stabilized asymptotically in a short time by the Hamiltonian structure-based feedback control law.

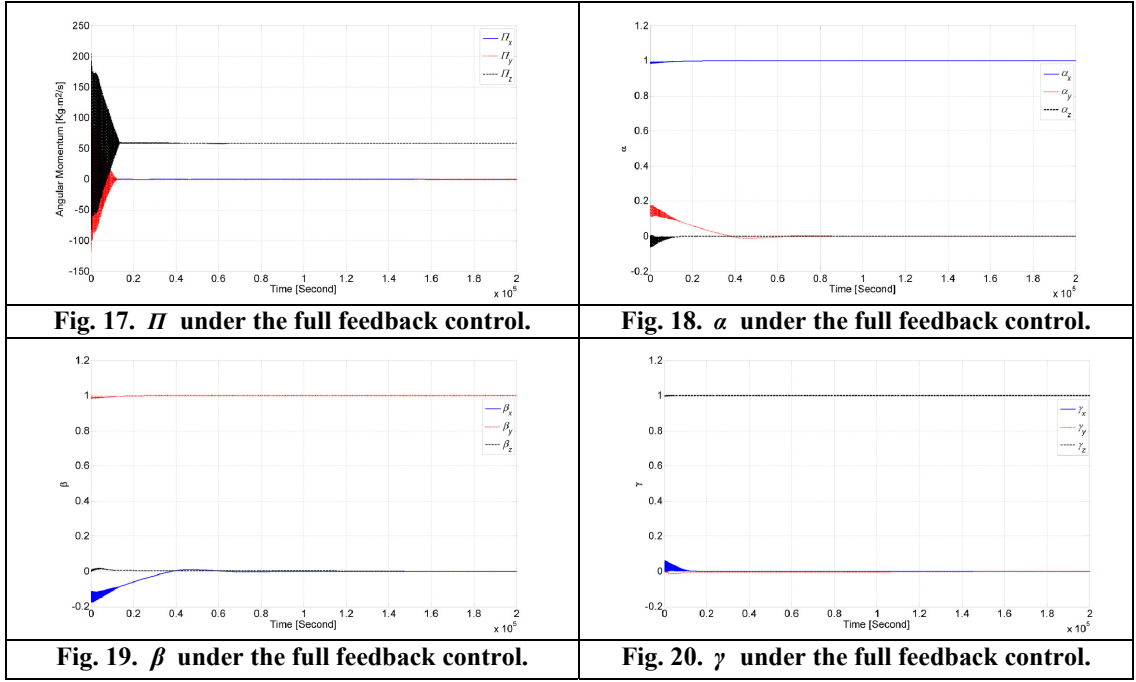

| Fig. 17. $\Pi$ under the full feedback control. | Fig. 18. $\alpha$ under the full feedback control. |
|---|---|
| Fig. 19. $\beta$ under the full feedback control. | Fig. 20. $\gamma$ under the full feedback control. |

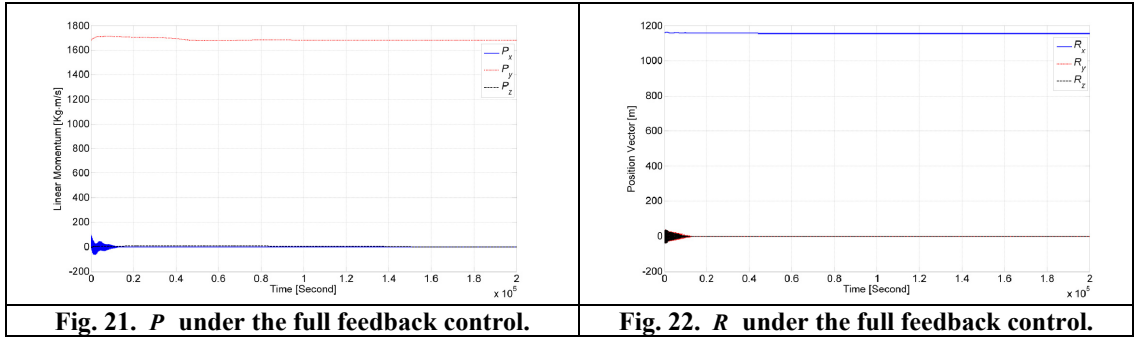

| Fig. 21. $P$ under the full feedback control. | Fig. 22. $R$ under the full feedback control. |
|---|---|

The energy $F_C(z)$ is shown by Fig. 23. We can see that the energy of the system converge to the value at the relative equilibrium under the energy dissipation control.

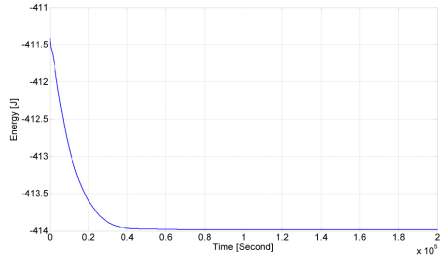

**Figure 23. Energy $F_C(z)$ under the full feedback control.**

The control torque and force acting on the spacecraft are shown by Figs. 24 and 25 respectively. We can find that the control torque and force are both converging to the zero, as the motion of the system is converging asymptotically to the relative equilibrium. The magnitudes of the control torque and force are very small, which means that the proposed Hamiltonian structure-based feedback control law is feasible and highly effective. The control torque and force can be provided by the attitude control system and low-thrust engines onboard the spacecraft respectively.

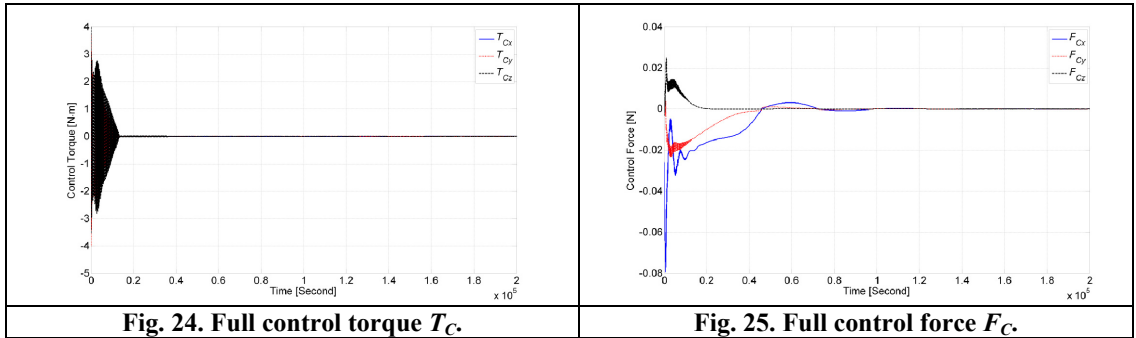

| Fig. 24. Full control torque $T_C$. | Fig. 25. Full control force $F_C$. |
|---|---|

Notice that both the control torque and force are expressed in the body-fixed frame of the spacecraft. Then, it is very convenient for the low-thrust engines fixed on the spacecraft to generate the required control force without a coordinate transformation.

The main advantage of the proposed hovering control law is that it is very simple and needs little computation to implement. This point is extremely important for the close proximity-operations near an asteroid, which needs a high level autonomy to overcome the complex environment and the long time delay in the communication from Earth. The simple form of the hovering control law is attributed to the full utilization of the non-canonical Hamiltonian structure in the control design.

## CONCLUSION

A Hamiltonian structure-based feedback control law has been proposed to stabilize the relative equilibria to achieve the orbit-attitude hovering near an asteroid. In this hovering model, both the position and attitude of the spacecraft are kept to be stationary in the asteroid body-fixed frame. This orbit-attitude hovering at the natural relative equilibria is discussed in the framework of the full spacecraft dynamics, in which the spacecraft is modeled as a rigid body with the gravitational orbit-attitude coupling. The feedback control law is consisted of two parts: potential shaping and energy dissipation. The potential shaping feedback has been constructed so that the relative equilibrium is to be a minimum of the modified Hamiltonian on the invariant manifold. Using the energy-Casimir method, we have proved that the unstable relative equilibrium can always be stabilized in the Lyapunov sense by the potential shaping with sufficiently large feedback gains. Then the energy dissipation leads the motion to converge asymptotically to the minimum of the modified Hamiltonian on the invariant manifold, i.e., the relative equilibrium.

The feasibility of both the potential shaping and the full feedback control law with energy dissipation have been verified through numerical simulations in the case of a spacecraft orbiting around a small asteroid. Simulation results have shown that the relative equilibrium can be successfully stabilized by the potential shaping in the Lyapunov sense. By introducing energy dissipation, the relative equilibrium can be stabilized asymptotically by the full feedback control law.

The simulation results have also shown that the magnitudes of the control torque and force are very small. The control torque and force are both expressed in the body-fixed frame of the spacecraft. It is very convenient for the low-thrust engines fixed on the spacecraft to generate the required control force without a coordinate transformation.

The main advantage of the proposed hovering control law is that it is very simple and needs little computation to implement. This is very suitable for the close proximity-operations near an asteroid, which needs a high level autonomy to overcome the complex environment and the long time delay in the communication from Earth. This simple form is attributed to the full utilization of dynamical behaviors of the system, i.e., the non-canonical Hamiltonian structure.

## ACKNOWLEDGMENTS

This work is supported by the Innovation Foundation of BUAA for PhD Graduates.

## REFERENCES

[1] Lara, M., and Scheeres, D. J.: Stability Bounds for Three-Dimensional Motion Close to Asteroids. Journal of the Astronautical Sciences, **50**(4), 389–409 (2002)

[2] Scheeres, D.J.: Orbit mechanics about asteroids and comets. J. Guid. Control Dyn. 35(3), 987–997 (2012)

[3] Scheeres, D.J.: Orbit mechanics about small bodies. Acta Astronaut. 72, 1–14 (2012)

[4] Scheeres, D. J.: Stability of hovering orbits around small bodies. Spaceflight Mechanics 1999, Advances in the Astronautical Sciences, Vol. 102, Pt. 2, Univelt, San Diego, CA, 1999, pp. 855–875.

[5] Sawai, S., Scheeres, D. J., and Broschart, S. B.: Control of hovering spacecraft using altimetry. Journal of Guidance, Control, and Dynamics, Vol. 25, No. 4, 2002, pp. 786–795.

[6] Broschart, S. B., and Scheeres, D. J.: Control of hovering spacecraft near small-bodies: application to asteroid 25143 Itokawa. Journal of Guidance, Control, and Dynamics, Vol. 28, No. 2, 2005, pp. 343–354.

[7] Broschart, S. B., and Scheeres, D. J.: Boundedness of spacecraft hovering under dead-band control in time-invariant systems. Journal of Guidance, Control, and Dynamics, Vol. 30, No. 2, 2007, pp. 601–610.

[8] Nazari, M., Wausony, R., Critzy, T., Butcher, E. A., and Scheeres, D. J.: Observer-based body-frame hovering control over a tumbling asteroid. The 2013 AAS/AIAA Astrodynamics Specialist Conference, AAS 13-820, Hilton Head, South Carolina, August 11–15, 2013.

[9] Koon, W.-S., Marsden, J.E., Ross, S.D., Lo, M., Scheeres, D.J.: Geometric mechanics and the dynamics of asteroid pairs. Ann. N. Y. Acad. Sci. **1017**, 11–38 (2004)

[10] Scheeres, D.J.: Spacecraft at small NEO. arXiv: physics/0608158v1 (2006)

[11] Wang, Y., Xu, S.: Gravitational orbit-rotation coupling of a rigid satellite around a spheroid planet. J. Aerosp. Eng. 27(1), 140–150 (2014)

[12] Lee, D., Sanyal, A. K., Butcher, E. A., and Scheeres, D. J.: Spacecraft hovering control for body-fixed hovering over a uniformly rotating asteroid using geometric mechanics. The 2013 AAS/AIAA Astrodynamics Specialist Conference, AAS 13-821, Hilton Head, South Carolina, August 11–15, 2013.

[13] Wang, Y., Xu, S.: Symmetry, reduction and relative equilibria of a rigid body in the $J_2$ problem. Adv. Space Res. 51(7), 1096–1109 (2013)

[14] Wang, Y., Xu, S.: Stability of the classical type of relative equilibria of a rigid body in the $J_2$ problem. Astrophys. Space Sci. 346(2), 443-461 (2013)

[15] Wang, Y., Xu, S.: Linear stability of the relative equilibria of a spacecraft around an asteroid. 64th International Astronautical Congress, IAC-13-C1.9.5, Beijing, China, Sept. 23–27, 2013 (2013)

[16] Wang, Y., Xu, S., Tang, L.: On the existence of the relative equilibria of a rigid body in the $J_2$ problem. Astrophys. Space Sci. doi: 10.1007/s10509-013-1542-y (in press, 2013)

[17] Wang, Y., Xu, S.: Hamiltonian structures of dynamics of a gyrostat in a gravitational field. Nonlinear Dyn. 70(1), 231–247 (2012)

[18] Beck, J.A., Hall, C.D.: Relative equilibria of a rigid satellite in a circular Keplerian orbit. J. Astronaut. Sci. 40(3), 215–247 (1998)

[19] Wang, Y., Xu, S.: Gravity gradient torque of spacecraft orbiting asteroids. Aircr. Eng. Aerosp. Tec. 85(1), 72–81 (2013)

[20] Marsden, J.E., Ratiu, T.S.: Introduction to Mechanics and Symmetry, TAM Series 17, Springer Verlag, New York (1999)

[21] Wang, Y., Xu, S.: On the nonlinear stability of relative equilibria of the full spacecraft dynamics around an asteroid. Nonlinear Dynamics (2014) DOI 10.1007/s11071-013-1203-2